\documentclass[runningheads]{llncs}

\usepackage[T1,T2A]{fontenc}
\usepackage[utf8]{inputenc}
\usepackage{graphicx}
\usepackage{color}
\usepackage{cite}
\usepackage[hidelinks]{hyperref}
\usepackage{multirow}
\usepackage{subcaption}

\usepackage[acronym, shortcuts]{glossaries-extra}

\glssetcategoryattribute{acronym}{nohyper}{true}
\setabbreviationstyle[acronym]{long-short}

\newacronym{ai}{AI}{Artificial Intelligence}
\newacronym{dl}{DL}{Deep Learning}
\newacronym{gdpr}{GDPR}{General Data Protection Regulation}
\newacronym{pii}{PII}{Personally Identifiable Information}
\newacronym{ml}{ML}{Machine Learning}
\newacronym{tab}{TAB}{Text Anonymization Benchmark}
\newacronym{lstm}{LSTM}{Long Short-Term Memory}
\newacronym{ner}{NER}{Named Entity Recognition}
\newacronym{nlp}{NLP}{Natural Language Processing}
\newacronym{echr}{ECHR}{European Court of Human Rights}
\newacronym{smote}{SMOTE}{Synthetic Minority Over-sampling Technique}
\newacronym{dt}{DT}{Decision Tree}
\newacronym{rf}{RF}{Random Forest}
\newacronym{svm}{SVM}{Support Vector Machine}
\newacronym{xgboost}{XGBoost}{eXtreme Gradient Boosting}
\newacronym{dnn}{DNN}{Deep Neural Network}
\newacronym{relu}{ReLU}{Rectified Linear Unit}
\newacronym{cnn}{CNN}{Convolutional Neural Network}
\newacronym{ocr}{OCR}{Optical Character Recognition}

\begin{document}

\title{RedactBuster: Entity Type Recognition from Redacted Documents}

\titlerunning{RedactBuster}

\author{Mirco~Beltrame\inst{1} \and
Mauro~Conti\inst{1,2} \and
Pierpaolo~Guglielmin\inst{1} \and
Francesco~Marchiori\inst{1} \and
Gabriele~Orazi\inst{1,3}
}

\institute{University of Padua, Department of Mathematics \and
Delft~University~of~Technology\\Faculty of Electrical Engineering, Mathematics and Computer Science \and
FDM Business Services \\
\email{\{mirco.beltrame.1, pierpaolo.guglielmin\}@studenti.unipd.it}\\
\email{\{gabriele.orazi, francesco.marchiori.4\}@phd.unipd.it}\\
\email{mauro.conti@unipd.it}}

\authorrunning{M. Beltrame et al.}

\maketitle

\begin{abstract}
The widespread exchange of digital documents in various domains has resulted in abundant private information being shared.
This proliferation necessitates redaction techniques to protect sensitive content and user privacy.
While numerous redaction methods exist, their effectiveness varies, with some proving more robust than others.
As such, the literature proposes several deanonymization techniques, raising awareness of potential privacy threats.
However, while none of these methods are successful against the most effective redaction techniques, these attacks only focus on the anonymized tokens and ignore the sentence context.

In this paper, we propose \textbf{RedactBuster}, the first deanonymization model using sentence context to perform Named Entity Recognition on reacted text.
Our methodology leverages fine-tuned state-of-the-art Transformers and Deep Learning models to determine the anonymized entity types in a document.
We test RedactBuster against the most effective redaction technique and evaluate it using the publicly available Text Anonymization Benchmark (TAB).
Our results show accuracy values up to 0.985 regardless of the document nature or entity type.
In raising awareness of this privacy issue, we propose a countermeasure we call \textit{character evasion} that helps strengthen the secrecy of sensitive information.
Furthermore, we make our model and testbed open-source to aid researchers and practitioners in evaluating the resilience of novel redaction techniques and enhancing document privacy.

\end{abstract}
\section{Introduction}
\label{sec:introduction}

Increasing document digitalization efforts have been adopted in several domains, such as the corporate sector, healthcare, and government~\cite{zhang2020research}.
These procedures allow for streamlined workflows, improving processing times and reducing reliance on manual manipulation.
As a side effect, the amount of data exchanged has also seen a significant rise~\cite{nabbosa2020societal}.
To justify the magnitude of the phenomenon, in 2022 and in Italy alone, the car and motorcycle insurance market generated more than 39 million documents between contracts and claims for a market of about 22.6 billion euro~\cite{ivass2022report}.
Each of these 39 billion documents contains sensitive information that needs to be protected.
While the process of digitalization can improve efficiency and optimization in specific tasks thanks to the recent advancements of \ac{ai} and data-driven approaches, it can cause privacy concerns.
Indeed, the lack of obfuscation of sensitive content may lead to unfair profiling of certain individuals or discrimination of specific groups~\cite{hajian2015discrimination}.
For these reasons, \textit{anonymization} defined as ``the process of rendering personal data anonymous'' has been mandated for \ac{gdpr} compliance in the EU~\cite{EuropeanParliament2016a}.

Several anonymization techniques have been proposed, given the importance of protecting user privacy and complying with regulations.
In the context of privacy protection, \textit{redaction} refers to the process of selectively editing or obscuring sensitive information from documents to prevent unauthorized access or disclosure~\cite{bier2009rules}.
It allows for removing only specific parts of a sentence while preserving the overall content of the text.
In the Italian case, document redaction is required for all public administrations.
The Personal Data Protection Guarantor (GPDP) obliges public organizations to publish how public resources were used by anonymizing \ac{pii} that refers to individuals.
Some examples of redaction techniques are shown in Fig.~\ref{fig:redaction}.
One redaction technique blurs a specific text part to make it unreadable (Fig.~\ref{subfig:blur}).
Another considers specific parts of the document as images and pixelates them by reducing their image quality (Fig.~\ref{subfig:pix}).
Blackout or whiteout redaction techniques are more effective, as they discard entirely the content to be anonymized and substitute it with black or white boxes (Fig.~\ref{subfig:black}).

\def\halfwidth{0.496085}
\begin{figure}[!htpb]
  \centering
  \begin{subfigure}{\halfwidth\textwidth}
     \centering
     \includegraphics[width=\textwidth]{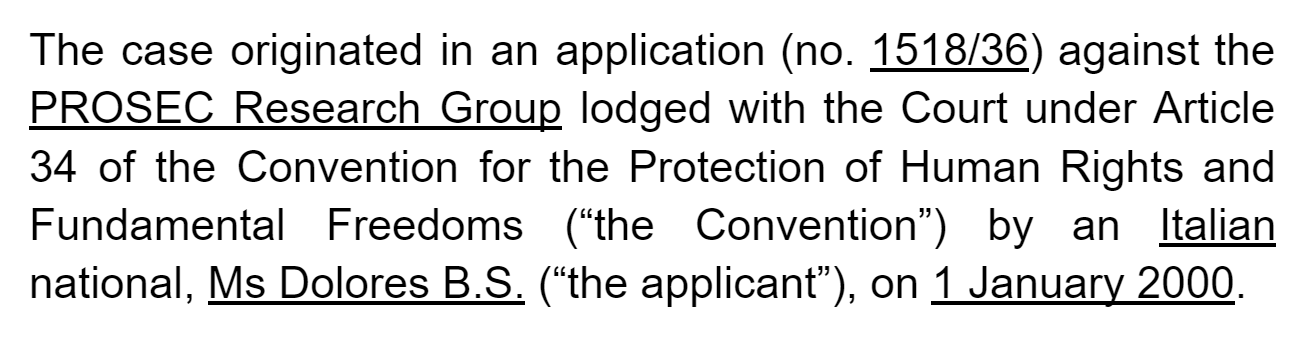}
     \caption{Plain text.}
     \label{subfig:unredacted}
  \end{subfigure}
  \begin{subfigure}{\halfwidth\textwidth}
     \centering
     \includegraphics[width=\textwidth]{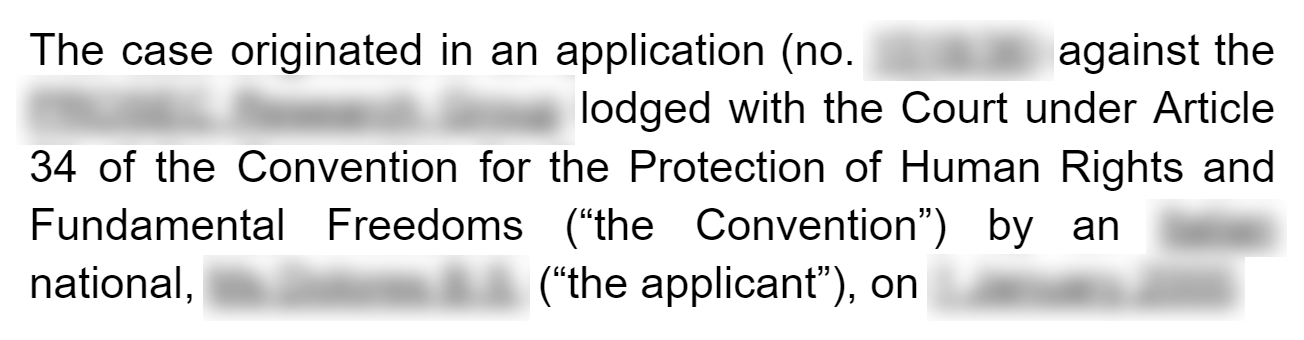}
     \caption{Blurring.}
     \label{subfig:blur}
  \end{subfigure}
  \begin{subfigure}{\halfwidth\textwidth}
     \centering
     \includegraphics[width=\textwidth]{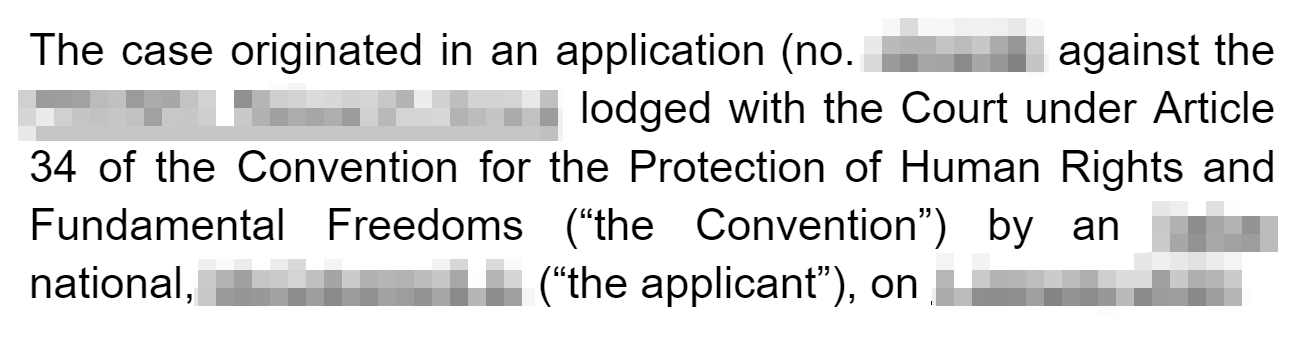}
     \caption{Pixelation.}
     \label{subfig:pix}
  \end{subfigure}
  \begin{subfigure}{\halfwidth\textwidth}
     \centering
     \includegraphics[width=\textwidth]{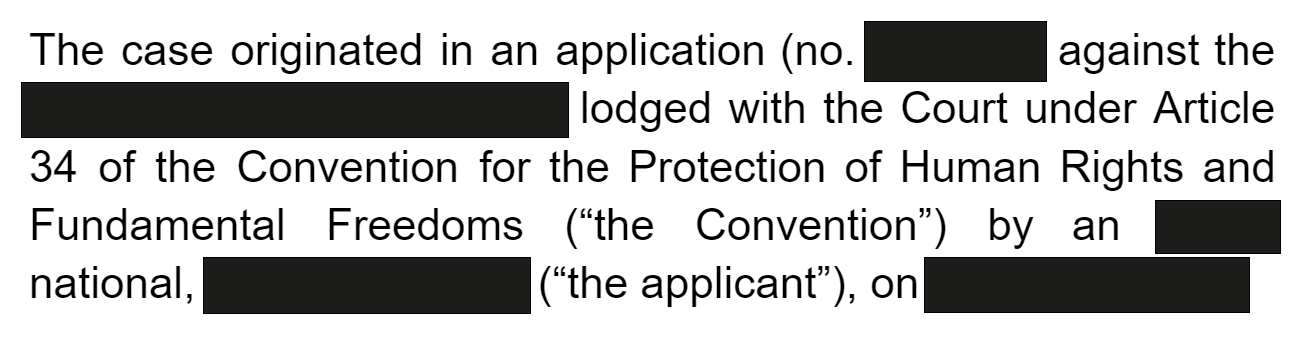}
     \caption{Blackout.}
     \label{subfig:black}
  \end{subfigure}
  \caption{Examples of redaction techniques.}
  \label{fig:redaction}
\end{figure}

While redaction techniques are commonly used in many domains, their security in protecting user privacy is uncertain.
Indeed, several works in the literature propose attacks for determining redacted content in sensitive documents.
For example, blurred and pixelated content can be unmasked with high accuracy~\cite{hill2016effectiveness}.
For these reasons, other redaction techniques, such as blackout and whiteout, should be preferable.
While the effectiveness of other redaction techniques has not yet been discussed, no attacks in the literature have been proposed.
Furthermore, the existing attacks focus exclusively on the redacted tokens.
Since with more effective anonymization techniques the sensitive content is erased from the document, this strategy is not effective anymore.

\paragraph{Contribution.}
In this paper, we present \textbf{RedactBuster}, the first document deanonymization attack against the most effective redaction technique.
{We open-source and evaluate our framework on the most comprehensive dataset, obtaining results of up to 0.985.
We also discuss the implementation of several countermeasures to our attack and propose a novel technique to enhance redaction efforts.
Our contributions can be summarized as follows.

\begin{itemize}
    \item We present \textbf{RedactBuster}, the first deanonymization attack against the most effective redaction technique.
    Our methodology leverages state-of-the-art \ac{ml} and \ac{dl} models for computing sentence embeddings and performing classification.
    \item We evaluate our framework on \ac{tab}, the most extensive dataset publicly available on document redaction~\cite{pilan2022text}.
    This dataset contains 1268 court cases in English, which have been properly annotated, labeled, and redacted.
    Our results show baseline accuracy values of 0.958, which can reach 0.985 by fine-tuning the embedding model on our specific task.
    We test several \ac{ml} and \ac{dl} models, comprehensively evaluating different architectures' capabilities and limitations.
    \item We propose \textit{character evasion} as an effective countermeasure against our attack and test them against our model.
    This technique involves the exchange of specific homoglyphs, allowing users to read and process the document effortlessly but preventing malicious parties from extracting the redacted entity types.
    We also demonstrate the effectiveness of our countermeasure on the dataset and show how swapping only five characters can decrease the attack success rate to 0.195.
    \item We open-source our framework at: \url{https://anonymous.4open.science/r/RedactBuster-1518}.
\end{itemize}

\paragraph{Organization.}
Our paper is organized as follows.
In Section~\ref{sec:related}, we discuss related works on anonymization and attacks on redaction techniques.
Section~\ref{sec:systemthreat} presents our system and threat model, while technical details on the methodology are provided in Section~\ref{sec:methodology}.
In Section~\ref{sec:evaluation}, we evaluate our framework, and we propose countermeasures in Section~\ref{sec:discussion}.
Finally, Section~\ref{sec:conclusions} concludes our work.
\section{Related works}
\label{sec:related}

The importance of redaction for protecting documents containing sensitive data or \ac{pii} does not have a widely shared standard among parties to date.
In June 2023, a litigation between Microsoft and the Federal Trade Commission demonstrated that \textit{(i)} documents can still be redacted by hand using a Sharpie marker and \textit{(ii)} redacted document represents a threat to organizations because of possible and expensive data leaks\footnote{\url{https://www.theverge.com/2023/6/28/23777298/sony-ftc-microsoft\\-confidential-documents-marker-pen-scanner-oops}}.
This event highlights how sensitive information circulating through documents is still subject to human error today. 
The unstructured nature of a written document makes it impossible to apply analytical methods to arrive at a specific result.
Despite the current tools to support this problem, manual validation is simultaneously the only solution to curb the deficiency and the source of possible issues.
Therefore, in this section, we describe which are the current solutions for document redaction, how \ac{pii} can be recognized within a text, and which unredaction techniques have been found to date.

\paragraph{Redaction techniques.}
Many studies and patents tackle the redaction phase per se, covering some parts of the text.
In \cite{kelly2013process}, the procedure is specifically crafted to work in synergy with Microsoft Office Word.
In \cite{matichuk2020redaction} a rule-set based automatic redaction system is presented, while \cite{cottrille2011selective} proposes a dynamic redaction based on tags specified by the user. Instead of covering, random generation of sensitive characters and numbers has also been patented~\cite{mane2023method}. 
A similar approach is presented in \cite{bendersky2017information}, which includes a redaction and storage system for redacted documents.
Another valuable approach is presented in \cite{ramos2020document}, which implies the concept of blockchain to preserve the security of an already-redacted document.
Unfortunately, such studies neglect the core of the redaction process, which is identifying the sensitive characters of a text that must be covered.

\paragraph{\ac{ner}.}
\ac{ner} is a crucial component of \ac{nlp}.
It identifies predefined categories of objects in text, such as names of individuals, organizations, locations, expressions of time, and quantities.
In \cite{chiu2016named}, the authors present a novel approach in the field by employing bidirectional \ac{lstm} and \ac{cnn}~architecture.
The insight of the latter work influenced the creation of the well-known BERT\cite{devlin2018bert}, a milestone on which much work in \ac{nlp} is still based today.
BERT has also proven to be highly effective in developing techniques for \ac{ner}~\cite{liu2019roberta, luoma2020exploring}.
Nevertheless, BERT models are also adopted as hyper-specialized models for peculiar domains such as \cite{zhao2021fine} materials and aerospace engineering~\cite{tikayat2023aerobert}.
The scientific community has shown approaches for automation of \ac{ner} and subsequent redaction of sensitive parts, albeit supervised by an end user~\cite{bier2009rules}.
On the same line, Microsoft developed and maintains an open source tool called PRESIDIO\cite{presidio}, which analyzes a text and, based on multiple recognizers, can detect and anonymize \ac{pii}.

\paragraph{Unredaction.}
Despite exciting developments in text redaction, studies are pointing out that some methods of information coverage are ineffective and vulnerable to possible data leaks.
In~\cite{bland2023story}, the authors assess standard PDF redaction tools vulnerability.
The study enlightens the possibility of recovering first and last names just because of subpixel-sized shifts of characters.  
Instead, the authors of~\cite{hill2016effectiveness} demonstrate the inadequacy of mosaicing, blurring, and pixelating methods as graphical coverage of sensitive characters.
To further demonstrate how dangerous such methods are, the security engineer Dan Petro released a proof-of-concept~\cite{githubGitHubBishopFoxunredacter}.
Although the outcome might be preferred to solid black rectangles, such aberrations represent an actual information leakage for a potential attacker.
\section{System and Threat Model}
\label{sec:systemthreat}

We now discuss the assumptions defining both the redaction system's functionality and potential attackers' capabilities.
In particular, Section~\ref{subsec:system} delves into the system model and the functionalities of a redaction pipeline in an adversary-free environment.
In Section~\ref{subsec:threat}, we analyze the attacker's knowledge of the target model in real-world scenarios.

\subsection{System Model}
\label{subsec:system}
Digital documents are commonly used online to share knowledge, sign contracts, or write reports.
Due to the nature of documents, it is expected to share such files while preserving the sensitive information.
\textit{Redaction} is the process that allows companies and organizations to identify and protect the most sensitive parts of a document by covering such portions of the text.
To date, no objective standard has been certified as the best methodology for digital document redaction. The security of such an operation falls in the hands of individuals or companies that may adopt more or less virtuous methods.

A general pipeline that organizations adopt is the one shown in Fig.~\ref{fig:redaction_pipeline} and includes the following main steps:
\begin{enumerate}
    \item \textit{Automated Entity Recognition.} The text in a document is fed into a \ac{ner} software, which produces an intermediate document version with highlighted entities that will potentially be redacted.
    Entities can be differentiated by using different highlighting colors depending on the kind of entity (e.g., proper name, date, location).
    If needed, entities that can or must be disclosed safely can be included in a whitelist.
    \item \textit{Human validation.} An operator takes charge of the intermediate document and manually reviews the highlighted entities to assess that there are no missed detections.
    Once the validation is complete, the final redacted version of the document can be generated. 
\end{enumerate}

\begin{figure}[!htpb]
    \centering
    \includegraphics[width=\columnwidth]{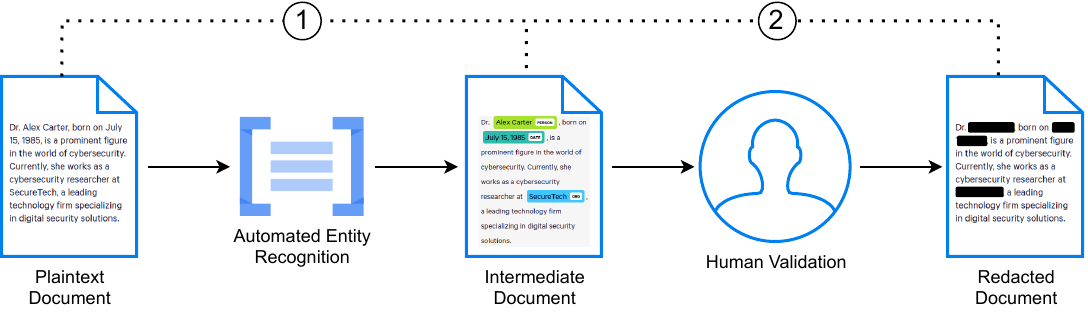}
    \caption{General pipeline for document redaction.}
    \label{fig:redaction_pipeline}
\end{figure}

Although redaction is a lossy process of the original digital document, many details can still be used to empower inference techniques for disclosure purposes.

\subsection{Threat Model}
\label{subsec:threat}
RedactBuster aims to predict which kinds of entities are hidden underneath a redacted text.
Even though working with unstructured text contained in documents is usually a challenge, our model wants to leverage the complexity of sentences to achieve its goal.

The attacker employing RedactBuster is interested in reconstructing the redacted text, ideally being able to rebuild the original unredacted document. In other words, they want to revert the pipeline described in Section~\ref{subsec:system}.
Documents are redacted to be shared with a broader audience that might not otherwise view the information in plain text.
For this reason, we assume that the attacker has access to a redacted document that, if deanonymized, could lead to real gain.
For instance, the attacker may want to access plaintext documents to extort, threaten, or ransom the victim.

\paragraph{Formal definition.}
We will now outline the context in which the attacker operates, considering four key criteria~\cite{biggio2018wild}: their \textit{goals}, \textit{knowledge}, \textit{capabilities}, and \textit{strategies}.

\begin{itemize}
    \item \textit{Goal:} The attacker wants to recover the original version of a document that has been redacted.
    \item \textit{Knowledge:} The attacker knows the area in which the document of interest falls.
    A list of examples includes fiscal, legal, or technical scope.
    \item \textit{Capability:} The attacker knows how and where to find redacted documents they aim to deanonymize.
    \item \textit{Strategy:} The attacker extracts the redacted text from the target document and processes it through the proposed model.
    With the entity predictions, they can leverage other kinds of attacks (such as social engineering attacks or data leaks) or knowledge bases to fully unredacted the target document.
\end{itemize}

\paragraph{Practical scenario.}
To accomplish the goal, the attacker of our threat model must progress through different stages: \textit{collection}, \textit{processing} and \textit{recomposition}: 

\begin{enumerate}
    \item \textit{Collection:} The attacker gains access to redacted documents. 
    Optionally, the attacker can use other publicly available datasets similar to the target document's scope to fine-tune the unredaction system.
    \item \textit{Processing:} After being extracted, the redacted text is processed by the proposed model to gain predictions on covered entity types.
    \item \textit{Recomposition:} Combining the results of the previous step and additional knowledge (personal/public domain or combined attacks), the attacker recovers the original document's plain text. 
\end{enumerate}

The main focus of this paper is mainly encapsulated in the \textit{processing} step.
The \textit{collection} step is dependent on the target that the attacker sets.
Similarly, in the \textit{recomposition} step, the attacker has a plethora of possibilities that are impossible to describe exhaustively and are, in fact, not primarily of interest to the purpose of the paper.
Some essential insights may already be contained within the redacted document: one example is the length of the redacted block, which can determine with a reasonable degree of approximation how many characters the hidden text consists of.

Along the same lines, the step of extracting text from the target document (typically a PDF file) is beyond the scope of this paper.
The documents are usually structured, and the text can be extracted even with a simple copy-paste from any file reader.
In cases where the document is unstructured because perhaps it is the result of scanning, \ac{ocr} can be involved.
This technology has been extensively examined by the academic community\cite{chen2021text, li2023trocr}~as well as being widely adopted in the consumer world.
In addition, there are \ac{ocr} open source projects such as Tesseract\footnote{\url{https://github.com/tesseract-ocr/tesseract}} or Apache Tika\footnote{\url{https://tika.apache.org/}}, both widely adopted by the developers.
\section{Methodology}
\label{sec:methodology}

We now detail our methodology and the techniques we use for unredaction.
In Section~\ref{subsec:dataset}, we provide the specifics of the dataset and how it generalizes real-world scenarios.
Section~\ref{subsec:preprocessing} discusses the data processing steps applied, constituting a crucial component of our contribution.
In Section~\ref{subsec:models}, we detail the used models and their hyper-parameters.
A complete overview of our framework is shown in Fig.~\ref{fig:pipeline}.

\begin{figure}[!htpb]
    \centering
    \includegraphics[width=\columnwidth]{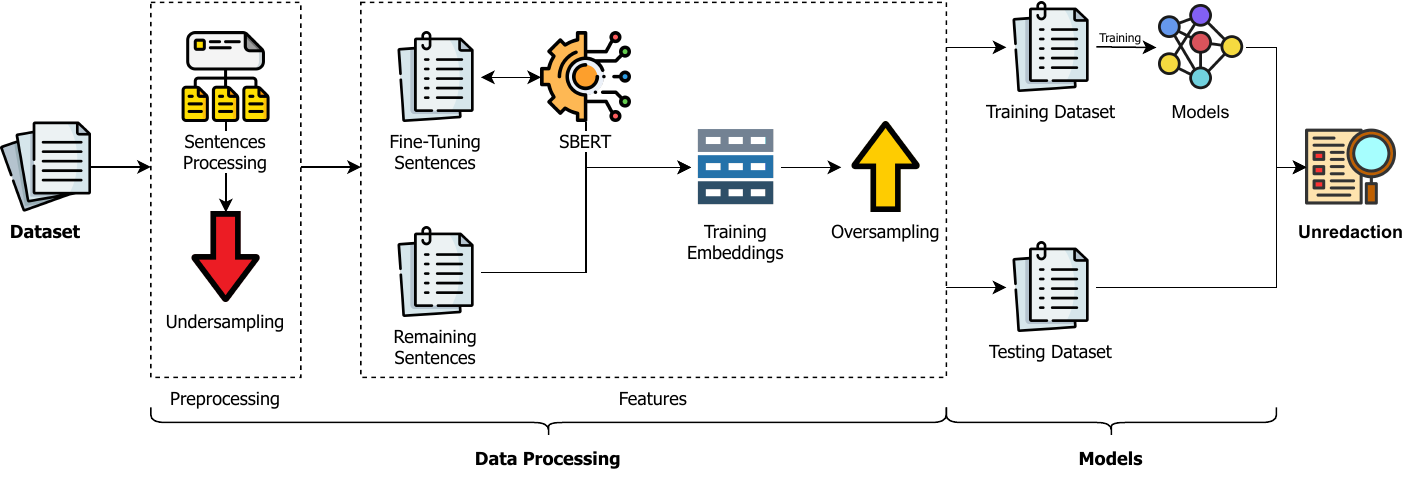}
    \caption{RedactBuster framework overview.}
    \label{fig:pipeline}
\end{figure}

\subsection{Dataset}
\label{subsec:dataset}

Publicly available datasets present several limitations that make their application to our study particularly challenging.
In particular, most datasets are restricted to clinical records and describe only a single individual~\cite{johnson2023mimic}.
Furthermore, many are related to public figures, include only short sentences, are composed of images, or have been subject to prior anonymization~\cite{papadopoulos2013impact}.

\paragraph{Text Anonymization Benchmark.}
For our study, we use the \ac{tab} dataset, an open-source annotated corpus consisting of 1268 English-language court cases from the \ac{echr}.
Twelve university students at the University of Oslo have annotated the documents of this corpus after being trained on their classification.
In particular, multiple users have annotated each document to cross-check their output.
In case of disagreements, conflicts were resolved among students based on the guidelines provided by the authors.
These guidelines indicate that each annotator should redact \textit{direct identifiers} and \textit{quasi-identifiers}.
The former refers to unique values given to an individual, e.g., full names, email addresses, and social security numbers.
The latter refers to publicly known information of an individual that doesn't enable identification when isolated, e.g., gender, nationality, postal code.
Each document is then distributed in JSON format and contains the following data.
\begin{itemize}
    \item Text of the court case used during the annotation.
    \item Document annotation.
    \item The target of the anonymization task.
    \item Whether another annotator revised the document.
\end{itemize}

\paragraph{Labels.}
The redacted text and its properties will constitute the starting point for extracting the features needed for classification.
Instead, the annotations represent our study's labels and ground truth.
In particular, each redacted entity belongs to one of eight classes: datetime (e.g., \textit{``15 December 1993''}), organization (e.g., \textit{``Court of Cassation''}), person (e.g., \textit{``Dr Price''}), demographic (e.g., \textit{``police officer''}), location (e.g., \textit{``London''}), miscellaneous (e.g., \textit{``1,053 sq. m''}), quantity (e.g., \textit{``2,000,000 Swedish kronor (SEK)''}), and code (e.g., \textit{``36619/03''}).

\subsection{Data Processing}
\label{subsec:preprocessing}

When dealing with classification tasks, features and their correlation are the most significant variables models can use to compute predictions.
For this reason, we first need to split the data for each annotation and define its ground truth.
Then, features must be extracted from the redacted sentences, and eventual biases in the dataset must be removed.

\paragraph{Preprocessing.}
The first step is to clean the text of each document and remove artifacts that might tamper with the classification process.
In this step, we handle end-of-line characters and particular abbreviations.
For instance, court texts in the dataset always include section titles (e.g., \textit{``THE FACTS''}), which are followed by a series of \texttt{\textbackslash n} characters.
We handle them by substituting them with dot characters and an appropriate number of spaces since their removal would make the redaction offset wrong.
We present more details of the textual-level modifications we employ in Appendix~\ref{app:preprocessing}.
Since each document contains several redacted entities, we separate them into single samples.
This facilitates classification, as \ac{ml} and \ac{dl} models predict one label for each sample.
Therefore, we split the text into sentences using NLTK PunktSentenceTokenizer~\cite{bird2006nltk}.
Each sentence is then associated with the calculated offset and entity type of the redactions that are present inside.
After this procedure, however, we might still encounter sentences containing multiple redacted entities.
For this reason, we duplicate the sentence and switch the redaction target until all confidential entities are redacted once.
This also serves as a data augmentation technique, as a single sentence can represent multiple labels depending on the redacted data.
An example is shown in Table~\ref{tab:duplicate}.

\begin{table}[!htpb]
  \centering
  \caption{Example of sentence splitting and redaction.}
  \label{tab:duplicate}
  \begin{tabular}{l|l|c|c}
    \hline
    \textbf{Sentence} & \textbf{Redaction} & \textbf{Offset} & \textbf{Type} \\
    \hline
    It happened on 19/10/2004.\: & It happened on **********.\: &  $\langle15, 24\rangle$ & \:\texttt{DATETIME}\: \\ \hline
    \multirow{2}{*}{Paolo was in Amsterdam.} & ***** was in Amsterdam. & $\langle0, 4\rangle$ & \texttt{PERSON} \\ \cline{2-4}
    & Paolo was in *********. & \:$\langle13, 21\rangle$\: & \texttt{LOC} \\
    \hline
  \end{tabular}
\end{table}

At this stage, the dataset comprises the redacted text, the redaction offset, and the redaction type.
However, we notice a substantial imbalance between the class distribution of the labels.
Indeed, the most frequent class is datetime with 34280 samples, while code comprises only 2781 samples.
This heavy skew in the data distribution can negatively impact a \ac{ml} model performance, as it would create bias in classification.
Therefore, we perform random undersampling on the data to balance the distribution.

\paragraph{Features.}
After the preprocessing steps, the training and testing dataset mostly contain samples in text format.
However, \ac{ml} and \ac{dl} traditionally require a constant number of features from each sample to compute a prediction.
For this reason, we use a widespread technique in the \ac{nlp} literature: computing embeddings~\cite{li2018word}.
Given the importance of sentence data in our dataset, we use SentenceTransformers (SBERT), a popular Python framework for generating text and image embeddings~\cite{reimers2019sentence}.
This Python package provides several transformer models with different sizes and performances.
We use \texttt{all-mpnet-base-v2} as it provides the best performances overall.
Indeed, in our threat model scenario, resources or computational overhead do not represent an obstacle, as document unredaction is not a time-constrained task.

SBERT models are generated and trained for general purposes and do not focus on specific domains.
As such, using an out-of-the-box model yields lower accuracy values with respect to a specifically trained model.
For this reason, we resort to fine-tuning.
With this procedure, we can retain the knowledge acquired by the model during the original training procedure and adapt it to our particular domain to increase performance.
In the case of SBERT transformers, fine-tuning is performed by comparing sentences with one another and providing a similarity score for the couple.
This score is the cosine similarity.
Given two vectors $A$ and $B$ of length $n$, this metric is defined as follows.
\begin{equation}
\label{eq:cos}
    \cos\left(\theta\right) = \frac{A \cdot B}{||A||\:||B||} = \frac{\sum\limits_{i=1}^{n}A_i B_i}{\sqrt{\sum\limits_{i=1}^{n}A_i^2} \sqrt{\sum\limits_{i=1}^{n}B_i^2}}.
\end{equation}
The output of this score is then normalized on a scale from 0 to 1, where higher values indicate higher similarity between the sentences.
Therefore, we extract a subset of the training dataset containing 250 samples for each label (i.e., 2000 total samples).
For each label, we fine-tune the model as follows: \textit{(i)} we group sentences in pairs (i.e., 125 for each label) and provide a similarity score of 0.8, and \textit{(ii)} we randomly pick a sentence from that label and a sentence from another random label (until obtaining 125 couples) and provide a similarity score of 0.2.
We select those values since we want our model to comprehend the similarity of the sentence context around the redaction without losing their individual properties.
Indeed, since embeddings are created from redacted sentences, our transformer model can only process the remaining sentence tokens.
Therefore, phrase context provides the knowledge the classifier requires to compute predictions.
It is worth noting that the 2000 samples used for fine-tuning the model are then discarded from the dataset.
We do this to ensure that our transformer model and the classifier are presented with new data and to prevent overfitting.

After fine-tuning, our transformer generates embeddings from the remaining training sentences (i.e., 2531 for each label).
The embedding size is determined by the transformer model used for computation, which, in our case, is 768.
We now oversample the data at the embedding level to compensate for the undersampling performed during preprocessing.
This is a widespread data augmentation technique used in \ac{ml} literature, as it can compensate for the possible lack of data in a dataset and allow the classifier to train on more samples~\cite{fernandez2018smote}.
It is also worth noting that, in our scenario, this procedure can be applied only after the embedding computation, as oversampling textual data can provide several artifacts that can instead decrease the classifier performance.
In particular, we use \ac{smote} as oversampling technique~\cite{chawla2002smote}.
This technique selectively generates synthetic samples through a nearest-neighbors approach for the minority class (which, in our case, are all the classes) by interpolating between existing instances.
We generate samples for each label until we obtain 3500 total samples for each class.
An overview of the data balancing process is shown in Appendix~\ref{app:balancing}.
We then split our dataset in training and testing respective size percentages of 85\% and 15\%.

\subsection{Models}
\label{subsec:models}

After dealing with the dataset and its feature extraction, we design a classifier for the unredaction task.
We present a comprehensive study of different state-of-the-art models by considering three \ac{ml} models and two \ac{dl} models.
For the \ac{ml} models, we perform GridSearch on different hyper-parameters configurations to find the best values for them.
In particular, we perform a 5-fold cross-validation, removing the need to include a validation set in the train and test dataset split.
GridSearch is performed on a balanced subset of the original dataset ($\sim$ 10000 samples).
Then, the found hyper-parameters are applied to the model to train on the whole training dataset.
\ac{dl} architectures are instead validated manually due to the need to test multiple architectures and layers.
Furthermore, being GPU-accelerated, they are significantly faster with respect to the \ac{ml} executed on the CPU.

\paragraph{ML Models.}
The first model we test is the \ac{rf}~\cite{breiman2001random}, an ensemble learning method that constructs multiple \ac{dt} during training.
Each tree is trained on a random subset of the dataset (bagging), and a random subset of features is considered for each split, which adds randomness and reduces overfitting.
Hyper-parameter search is performed on the number of estimators, the criterion, and the max depth.
The first variable determines the number of trees in the forest (tested with 150, 200, and 300).
The second represents the function of measuring the quality of a split (tested with gini, entropy, and log loss).
The third describes the maximum depth of each \ac{dt} in the forest, thus controlling the complexity of the model (tested with 3, 5, or unlimited).

Another widespread model used in the literature is \ac{svm}~\cite{cortes1995support}.
It works by finding the hyperplane that best separates classes in a high-dimensional space.
For nonlinear classification tasks, \ac{svm} uses a kernel trick to map the original features into a higher-dimensional space where a separating hyperplane exists.
Hyper-parameter search is performed on the kernel and the C parameter.
The first variable specifies the kernel type for the algorithm (tested with polynomial and radial basis function).
In the case of the polynomial kernel, we also search for its optimal degree (tested with 3 and 4).
The second represents the regularization parameter that controls the trade-off between maximizing the margin and minimizing the classification error (tested with 3, 5, and 7).
We also use bagging (bootstrap aggregation) to make the training process faster and reduce overfitting.

Finally, we test \ac{xgboost}~\cite{chen2016xgboost}.
This model implements gradient-boosting algorithms designed for speed and performance.
It builds multiple decision trees iteratively and tries to correct the errors of the previous models.
It uses a gradient descent algorithm to minimize the loss function when adding new models.
Hyper-parameter search is performed on the learning rate, the L2 regularization, and the max depth.
The first parameter prevents overfitting by shrinking the feature weights to make the boosting process more conservative (tested with 0.01, 0.1, and 0.3).
The second variable is a regularization term on weights that penalizes the complexity of the model, preventing overfitting (tested with 0, 0.125, and 0.25).
The third represents the maximum depth of each tree (tested with 3, 5, or unlimited).

\paragraph{DL Models.}
The first \ac{dl} model we test is a feedforward \ac{dnn} composed of multiple layers of neurons.
Our specific implementation consists of five fully connected layers.
The input layer consists of 768 neurons, corresponding to the data's input features.
We include four hidden layers with 512, 256, 128, and 64 neurons, respectively.
Each hidden layer applies a linear transformation followed by a \ac{relu} activation function.
Finally, the output layer consists of 8 neurons, representing the final output classes.
The model is trained with cross-entropy as the loss function and uses Adam as an optimizer with a learning rate of 0.00005 for 200 epochs with a batch size of 100.

The other model architecture we use is the \ac{cnn}.
It is a type of neural network particularly effective for processing grid-like data, such as images or, in this case, one-dimensional sequences.
\acp{cnn} utilize convolutional layers to automatically and adaptively learn spatial hierarchies of features from the input data.
The reason why we select this specific architecture is for its capability to detect patterns in data samples.
Indeed, in text data, embedding captures semantic information and relationships between words or tokens~\cite{xu2016text}.
The convolutional layers can effectively process these to detect specific sequences at different abstract levels, such as character-level, word-level, and higher-level semantic features.
In our implementation, the first two layers are convolutional.
The first layer applies 16 filters of size 16 to the input sequence, producing feature maps.
The second layer applies 16 filters of size 16 with a stride of 2, reducing the spatial dimensions.
After each convolutional layer, a max pooling operation is applied with a kernel size of 8 and a stride of 2, which reduces the spatial dimensions while retaining significant features.
Following the convolutional layers, there are four fully connected layers with 1376, 688, 344, and 172 neurons, respectively.
These layers further process the extracted features.
Finally, the output layer consists of 8 neurons, representing the final output classes.
The model is trained with cross-entropy as the loss function and uses Adam as an optimizer with a learning rate of 0.0001 for 200 epochs with a batch size of 100.
\section{Evaluation}
\label{sec:evaluation}

We now evaluate the entity-type recognition capabilities of our proposed methodology.
First, in Section~\ref{subsec:metrics}, we disclose the metrics we use for the evaluation and their definition.
We then provide in Section~\ref{subsec:baseline} a baseline evaluation of or models in scenarios in which the Transformer model has not been fine-tuned on the corpus.
We finally show the effectiveness of our fine-tuning procedure in Section~\ref{subsec:ft}.

\subsection{Metric}
\label{subsec:metrics}

From a \ac{ml} perspective, our task is a multiclass classification task in which each label has the same importance.
This is different from a binary classification task, where there is usually a positive and negative class.
For this reason, in our scenario, it is not possible to define False Positives (FP), False Negatives (FN), and True Negatives (TN).
Instead, only two events can occur when the model predicts the redacted entity type.
\begin{itemize}
    \item \textit{True Positive (TP)}: the predicted entity type matches the original entity type.
    \item \textit{Misclassification (Err)}: the predicted entity type does not match the original entity type.
\end{itemize}
For this reason, we use accuracy as our evaluation metric, defined as the ratio of correctly classified instances (across all classes) to the total number of samples in the dataset.
\begin{equation}
    Accuracy = \frac{\sum\limits_{i=1}^8 TP_i}{\sum\limits_{i=1}^8 \left( TP_i + Err_i \right)} = \frac{\textrm{\# Correct Predictions}}{\textrm{\# Total Predictions}}.
\end{equation}

\subsection{Baseline}
\label{subsec:baseline}

We now present the evaluation of our models in baseline performance.
This implies that each model is trained on embeddings generated from a non-finetuned transformer model.
As such, after processing the document sentences as detailed in Section~\ref{subsec:preprocessing} and undersampling them based on the number of classes, we directly compute the embeddings with the out-of-the-box model provided by SBERT.
Since in this scenario the 2000 samples for fine-tuning the Transformer are not discarded, we reduce the oversampling percentage to reach the same number of total data samples.
Training and testing datasets are then generated with the same split percentages.
The results of this evaluation are shown in Table~\ref{tab:baseline}.
\ac{dl} models outperform all \ac{ml} models on the test dataset, with the \ac{dnn} model obtaining the best score.
It is also worth noting that \ac{ml} models are overfitting in most cases.
This can be observed by the high scores on the training dataset, which, however, are not representative of the models' capabilities on unforeseen data.
The \ac{ml} models' behavior in this task can be attributed to two main factors: \textit{(i)} the shallower architectures employed with respect to the \ac{dl} models, and \textit{(ii)} the variance and complexity of the embeddings.
Indeed, the \ac{dl} models reach high scores without overfitting, making their more complex architectures advantageous in this task.
Furthermore, high training scores but lower testing scores indicate that the models are memorizing the embeddings, which causes a loss of generalization capabilities.
This indicates that the vectorial representation of the sentences is too statistically varied, making smaller models unable to classify them.

\begin{table}[!htpb]
  \centering
  \caption{Baseline evaluation of the models.}
  \label{tab:baseline}
  \begin{tabular}{l|c|c}
    \hline
    \multirow{2}{*}{\textbf{Model}} & \multicolumn{2}{c}{\textbf{Accuracy}} \\ \cline{2-3}
    & Train & Test \\
    \hline
    \ac{rf}	& \:1.000\: & 0.697 \\
    \ac{svm} & 0.892 & 0.747 \\
    \ac{xgboost}\: & 0.970 & 0.714 \\
    \textbf{\ac{dnn}} & 0.979 & \:\textbf{0.958}\: \\
    \ac{cnn} & 0.981 & 0.924 \\
    \hline
  \end{tabular}
\end{table}

\subsection{Finetuning}
\label{subsec:ft}

To address the shortcomings of the baseline evaluations, we proposed using a fine-tuned transformer model for embedding computation.
In Table~\ref{tab:fintetuned}, we provide an evaluation of our models trained and tested on the fine-tuned dataset presented in Section~\ref{subsec:preprocessing}.
Most of the models present an increased score in both test and train datasets with respect to the baseline of Table~\ref{tab:baseline}.
Indeed, the best-performing model is now the \ac{cnn}.
Furthermore, we can also notice a significant improvement in the \ac{ml} models' scores.
This growth shows that fine-tuning the transformer model makes the embedding more homogeneous, allowing for using less complex models.
Regardless, the deeper architectures provided by \ac{dl} should be preferred, as they can provide consistently high scores and prevent overfitting.

\begin{table}[!htpb]
  \centering
  \caption{Evaluation of the models on fine-tuned embeddings.}
  \label{tab:fintetuned}
  \begin{tabular}{l|c|c}
    \hline
    \multirow{2}{*}{\textbf{Model}} & \multicolumn{2}{c}{\textbf{Accuracy}} \\ \cline{2-3}
    & Train & Test \\
    \hline
    \ac{rf}	& \:1.000\: & 0.823 \\
    \ac{svm} & 0.983 & 0.840 \\
    \ac{xgboost}\: & 0.940 & 0.866 \\
    \ac{dnn} & 0.946 & 0.945 \\
    \textbf{\ac{cnn}} & 0.986 & \:\textbf{0.985}\: \\
    \hline
  \end{tabular}
\end{table}
\section{Countermeasures}
\label{sec:discussion}

Given the high accuracies demonstrated by our evaluation in Section~\ref{sec:evaluation}, it is clear that sentence context can provide enough information for an attacker to infer the entity type of a redacted token.
This threat can constitute a critical privacy breach since, through the entity type, data linkage can be performed, which might lead to bias and discrimination.
Therefore, it is crucial to develop effective countermeasures depending on the field of application of the redaction.
For example, if sensitive documents are shared in PDF format, an option is to disable printing functionalities, thus preventing possible attackers from detecting and extracting textual information.
However, this can be counter-effective, as it also impedes legitimate parties from engaging with the document.
Unfortunately, this side effect also applies to other scenarios where text is extracted through \ac{ocr} systems.
For instance, when redacted documents are stored in paper format, their digitalization involves scanning and extracting textual data from retrieved images.
Thus, to prevent an attacker from being able to extract the text, adversarial attacks against \ac{ml}-based \ac{ocr} systems can be crafted~\cite{song2018fooling, xu2020machines}.
However, this also prevents legitimate parties from obtaining the document's contents.

\paragraph{Character Evasion.}
To address the shortcomings of traditional countermeasures against document unredaction, we propose a technique we call \textit{character evasion}.
This method involves substituting specific ASCII characters at test time to fool the entire unredaction pipeline.
Taking inspiration from adversarial and evasion techniques, the perturbations applied to the textual data are imperceptible to the human eye but are substantial enough to fool the \ac{ml} systems~\cite{goodfellow2014explaining}.
Indeed, one of our study's assumptions is that the attacker can access (or generate) a dataset of redacted documents.
As such, the attacker fine-tunes its Transformer model and trains the \ac{dl} models to process the text automatically.
Traditional evasion techniques would use the models' gradients to compute adversarial attacks at the embedding level to cause misclassification.
However, the attacker owns the unredaction pipeline; thus, legitimate parties cannot access the models' parameters and architectures.
Therefore, countermeasures must be applied at the document level to make the attacker misclassify redacted entity types.
In this case, the best strategy is to substitute characters in the text with graphically identical ones, which, however, are processed differently by models.
For example, each ``a'' character in the text can be swapped with the ``а'' character.
These characters are homoglyphs, i.e., they appear visually similar but are distinct characters with different Unicode code points.
In particular, the former is a Latin character (\texttt{U+0061}), while the latter is a Cyrillic character (\texttt{U+0430}).
We only swap the five most common letters in the English alphabet (for which an identical Cyrillic or Armenian character is available) and show them graphically with their Unicode code in Appendix~\ref{app:charev}.
This technique is particularly effective in cases where human readability is necessary while preserving the content's privacy.
Indeed, while humans can perfectly read the text, the unredaction attack is no longer effective.
An evaluation of this countermeasure is shown in Table~\ref{tab:evasion}.
When using character evasion, the accuracy of the models drops to values close to random guessing.
This behavior can be attributed to the Transformer model, which, by processing foreign characters at test time, cannot generate embeddings statistically close to the ones on which the \ac{dl} models are trained.
\ac{ml} models obtain similar results, with an average accuracy of 0.238.
Furthermore, we can also notice that fine-tuning the Transformer does not improve the results, making character evasion a robust countermeasure against our attack.

\begin{table}[!htpb]
  \centering
  \caption{Evaluation of the models with the \textit{character evasion} countermeasure.}
  \label{tab:evasion}
  \begin{tabular}{l|c|c}
    \hline
    \multirow{2}{*}{\textbf{Model\:}} & \multicolumn{2}{c}{\textbf{Accuracy}} \\ \cline{2-3}
    & \:Baseline\: & \:Fine-tuned\: \\
    \hline
    \ac{dnn} & 0.182 & 0.195 \\
    \ac{cnn} & 0.181 & 0.183 \\
    \hline
  \end{tabular}
\end{table}
\section{Conclusions}
\label{sec:conclusions}

With the increase in document digitalization and the consequent rise in the exchange of data, it is essential to ensure that sensitive content stays private.
For this reason, several redaction techniques have been proposed and used to mask private tokens inside textual documents.
With the recent discoveries on the insecurity of specific techniques such as blurring and pixelation, the complete removal of the token is often preferred as, in this way, the entity is erased from the document.
However, sentence context can still leak information on the entity type that has been redacted.

\paragraph{Contribution.}
This paper presented \textbf{RedactBuster}, the first entity type extraction attack on redacted tokens.
Our attack leverages state-of-the-art \ac{dl} models for the processing and classification of redacted sentences.
We evaluated our attack on a real-world dataset, obtaining an accuracy of 0.958.
Furthermore, we proposed an effective countermeasure called \textit{character evasion}, which can aid practitioners in defending against our discovered attack.

\paragraph{Future Works.}
For future endeavors, we aim to create a new dataset that can aid this line of research.
Indeed, while our dataset included several types of documents, more entity types and document styles can further solidify the contributions provided in this paper.
This opportunity also opens up the possibility of implementing \ac{ocr} system in the experimentations, thus simulating a complete document digitalization and redaction pipeline.
Access to a more realistic framework can aid us in detecting new vulnerabilities and developing more robust solutions to ensure user privacy.

\bibliographystyle{splncs04}
\bibliography{references}
\clearpage
\appendix
\section{Dataset}
\label{app:dataset}

In this Appendix section, we give more details on some of the processing steps that the text must go thought before being fed to the Transformer for embedding computation.

\subsection{Character Preprocessing}
\label{app:preprocessing}

After handling the section titles, we substitute all \texttt{\textbackslash n} characters with spaces, as some sentences might be contained in multiple lines.
This procedure also ensures that the overall number of characters in the text remains the same.
We also address possible abbreviations that are common in specific tokens.
For example, ``no.'' is often used for ``number'' (or ``nos.'' for ``numbers'').
These words cause the tokenizer to end a sentence after the dot character.
To fix this, we swap the dot character with an underscore.
Nevertheless, the tokenization process introduces several artifacts in the text, as it often removes spaces after a dot character at the end of a sentence.
This causes the sentences' offsets to be shifted with respect to the original text.
We treat sentences separately and compute the redaction offsets w.r.t. to the sentence start to solve this.
Finally, once we find the word to censor inside the sentence, we substitute it with the right amount of asterisk characters.

\subsection{Data Balancing}
\label{app:balancing}

To ensure that the models do not have any biases for classification, we balance the number of data samples for each class with different techniques.
An overview is shown in Table~\ref{tab:balance}.
First, we randomly undersample data to obtain a balanced distribution.
Then, we use a fixed amount of samples for each class for fine-tuning the Transformer model.
Finally, with the remaining dataset, we perform oversampling.

\begin{table}[!htpb]
  \centering
  \caption{Number of samples for each class at different stages of the unredaction pipeline.}
  \label{tab:balance}
  \begin{tabular}{l|c|c|c|c}
    \hline
    \textbf{Class} & \textbf{\:Dataset\:} & \textbf{\:Undersampling\:} & \textbf{\:Fine-Tuning\:} & \textbf{\:Oversampling\:} \\
    \hline
    DATETIME & 34280 & 2781 & 2531 & 3500 \\
    ORG & 28828 & 2781 & 2531 & 3500 \\
    PERSON & 13393 & 2781 & 2531 & 3500 \\
    DEM & 6325 & 2781 & 2531 & 3500 \\
    LOC & 6224 & 2781 & 2531 & 3500 \\
    MISC & 5169 & 2781 & 2531 & 3500 \\
    QUANTITY & 2963 & 2781 & 2531 & 3500 \\
    CODE & 2781 & 2781 & 2531 & 3500 \\
    \hline
    \textbf{Total} & 122963 & 22248 & 20248 & 28000 \\
    \hline
  \end{tabular}
\end{table}

\section{Character Evasion}
\label{app:charev}

In Table~\ref{tab:characters}, we show all the characters we use for our \textit{character evasion} technique and their respective Unicode code.

\begin{table}[!htpb]
  \centering
  \caption{List of character evasion subsitutions.}
  \label{tab:characters}
  \begin{tabular}{c|c||c|c}
    \hline
    \multicolumn{2}{c||}{\textbf{Original}} & \multicolumn{2}{c}{\textbf{Evasion}} \\ \hline
    \textbf{\:Char\:} & \textbf{\:Code\:} & \textbf{\:Char\:} & \textbf{\:Code\:} \\
    \hline
    a & \texttt{\:U+0061\:} & а & \texttt{\:U+0430\:} \\
    e & \texttt{\:U+0065\:} & е & \texttt{\:U+0435\:} \\
    i & \texttt{\:U+0069\:} & і & \texttt{\:U+0456\:} \\
    n & \texttt{\:U+006E\:} & n & \texttt{\:U+0578\:} \\
    o & \texttt{\:U+006F\:} & о & \texttt{\:U+043E\:} \\
    \hline
  \end{tabular}
\end{table}

\section{Hardware and Software Configuration}
\label{app:specs}

All experiments have been conducted on a workstation with the following configurations.
\begin{itemize}
    \item \textbf{CPU}: AMD Ryzen 5 3600X.
    \item \textbf{RAM}: 32 GB at 3200 MT/s
    \item \textbf{Operating System}: Ubuntu 20.04.4 LTS.
    \item \textbf{Software}: Python 3.8.10.
\end{itemize}
All \ac{ml} models are implemented with the \texttt{Scikit-learn} Python package, which does not natively support GPU acceleration.
Instead, \ac{dl} models are implemented with Pytorch 1.7.1.
Other Python packages and their related versions can be found in our repository's requirements file.\footnote{\url{https://anonymous.4open.science/r/RedactBuster-1518/requirements.txt}}

\end{document}